\documentclass[a4paper,11pt]{article}

\pdfoutput=1

\usepackage{jcappub} 
\usepackage{float}
\usepackage{mathtools}
\usepackage[T1]{fontenc}
\usepackage{subfig}
\usepackage{booktabs}
\title{\boldmath Fisher matrix for multiple tracers: the information in the cross-spectra}

\author[a]{L. Raul Abramo,}
\author[a]{Jo\~ao Vitor Dinarte Ferri,}
\author[a]{Ian Lucas Tashiro}
\affiliation[a]{\small Departamento de F\'{\i}sica Matem\'atica, Instituto de F\'{\i}sica, Universidade de S\~ao Paulo,\\ R.  do  Mat\~ao  1371,  05508-090,  S\~ao Paulo, SP, Brazil}

\emailAdd{abramo@if.usp.br}
\emailAdd{joao.vitor.ferri@usp.br}
\emailAdd{iantashiro@usp.br}

\abstract{We derive general expressions for the multi-tracer Fisher matrix, both assuming that the cross-spectra are constrained by the auto-spectra, and also allowing for independent degrees of freedom in the cross-spectra. 
We show that, just like the ratios of power spectra, the independent degrees of freedom of the cross-spectra are also not constrained by cosmic variance. Moreover, whereas the uncertainties in the ratios of power spectra decrease with the number density of the tracers as $\sim 1/\sqrt{\bar{n}}$, the uncertainties in the independent degrees of freedom of the cross-spectra decrease even faster, as $\sim 1/\bar{n}$. We also derive simple expressions for the optimal number of tracers in a survey.}


\begin{document}

\maketitle
\flushbottom

%%%%%%%%%%%%%%%%%%%%%%%%%%%%%
\section{Introduction}
\label{sec:introduction}
%%%%%%%%%%%%%%%%%%%%%%%%%%%%%

The new generation of galaxy surveys are mapping large swaths of the observable Universe using a variety of objects, and with increasing completeness \cite{Abbetal,Amendola:2012ys,Ivezic:2008fe,Benitez:2014ibt}.
These surveys allow us to infer cosmological information from the positions of galaxies and other tracers of large-scale structure mainly through measurements of their clustering.
In particular, when combining two or more tracers, we are able to measure not only their auto-spectra (or, equivalently, their auto-correlations), but also their cross-spectra.
And as we increase the diversity of tracers in our surveys, the relative importance of the cross-spectra grows as well: given $N$ tracers there are $N$ auto-spectra, and $N(N-1)/2$ cross-spectra.

A further advantage of multiple tracers is the fact that some physical parameters can be measured with an accuracy that is not limited by cosmic variance -- i.e., we are not necessarily constrained only by the survey volume, and can improve the measurements of some physical parameters by increasing the numbers of tracers in the same volume \cite{Seljak:2008xr,McDonald:2008sh}. As shown by \cite{abramo2013multitracer}, the independent degrees of freedom measured by galaxy surveys are split into two branches: on one hand, the total clustering of the survey (a single degree of freedom), which includes observables such as the shape of the matter power spectrum, is severely constrained by cosmic variance: no matter how many tracers we observe, we are limited by the volume of the survey, which imposes a lower bound to the uncertainties in that total clustering.
On the other hand, the {\em ratios} of power spectra of the tracers (or {\em relative} clusterings) are independent from the total clustering, as well as from each other, and their covariance is not limited by cosmic variance: it's always possible to beat down the noise in those variables by detecting larger numbers of tracers in any given survey volume.

However, in this argument it is often assumed that the auto-power spectra and the cross-spectra are manifestations of the same basic degrees of freedom: the biases, the matter growth rate, the amplitude and shape of the matter power spectrum, etc. 
In practice, this is equivalent to assuming that $P_{ij} = v_i v_j$, and taking these $N$ parameters $v_i$ as the fundamental degrees of freedom.
In fact, in linear theory the information in the cross-spectra is degenerate with that already contained in the auto-spectra -- for a review see, e.g., \cite{Desjacques2016}. 
A linear biasing model implies that the {\em observed} power spectra (the data covariance in Fourier space), in redshift space, is given by:
\begin{equation}
    \label{Eq:ObsPower}
    P_{ij} ( \vec{k} ) 
    = b_i b_j P_{m} ( \vec{k} )  + \frac{\delta_{ij}}{\bar{n}_i}
    \; \to \; (b_i + f \mu^2)(b_j + f \mu^2) P_m(k) + \frac{\delta_{ij}}{\bar{n}_i} \; ,
\end{equation}
where $b_i$ are the biases of the tracers $i=1,2,\ldots,N$, $f$ is the matter growth rate, $\mu = \hat{r}\cdot \hat{k}$ is the angle of the Fourier mode with the line of sight, and $\bar{n}_i$ are the number densities of the tracers. In the second part of the expression above we also used the flat-sky approximation to indicate explicitly the bias model -- but this assumption is irrelevant for the arguments presented in this paper.

Although the assumption that the degrees of freedom in the cross-spectra are contained in the auto-spectra is approximately correct in the linear regime, and assuming a deterministic biasing model, it misses many important features such as the one-halo term, exclusion effects, stochasticity, as well as deviations from the ideal Poisson shot noise model \cite{Baldauf:2013hka,Assassi2014,mirbabayi2015biased,Desjacques2016,Schmittfull:2018yuk}.

Extensions of the linear and deterministic biasing model motivated by perturbation theory introduce additional dependencies in the tracer density contrasts, typically of the form:
\begin{equation}
    \label{eq:deltai}
    \delta_i \to B_i \delta_m + \epsilon_i^P + \epsilon_i^S \; ,
\end{equation}
where $B_i=b_i + f \mu^2$, and
$\epsilon_i^P$ is a shot noise stochastic term that, under the assumption of Poisson statistics, obeys $\langle \epsilon_i^P \epsilon_j^P \rangle = \delta_{ij}/\bar{n}_i$. 
The last term in Eq. \eqref{eq:deltai}, $\epsilon_i^S$, collects
all additional stochastic terms and non-linearities (e.g., the dependencies on $\delta_m^2$).
A more general relation between the matter density field and the tracers, similar to Eq. \eqref{eq:deltai}, was in fact used by Gil-Mar\'{\i}n et al. \cite{2010MNRAS.407..772G} to model how the multi-tracer approach could improve measurements of the matter growth through RSDs.
In some cases, discarding these stochastic terms, especially in the cross-spectra, can lead to systematics in the measurements of parameters such as $f_{NL}$ \cite{Ginzburg}.
Recently, Mergulh\~ao et al. \cite{2021arXiv210811363M} showed that by splitting a halo population in two according to their mass  allows us to measure these bias parameters with higher precision, which implies that, in practice, the advantages of the multi-tracer approach can leverage the larger number of parameters that comes together with considering more tracers.

Let's assume, for simplicity, that shot noise is exactly Poissonian and is  uncorrelated with either $\delta_m$ or $\epsilon_i^S$.
The power spectrum corresponding to the pair $\langle \delta_i \delta_j \rangle$, after shot noise subtraction, takes the form:
\begin{equation}
    \label{eq:deltai2}
    \hat{P}_{ij} \to 
    B_i B_j P_{m}
    + \langle \epsilon_i^S \epsilon_j^S \rangle 
    \; ,
\end{equation}
where we assumed that the stochastic terms are uncorrelated with the density contrast.
Even if we assume the simplest stochastic model, $\langle \epsilon_i^S \epsilon_j^S \rangle = \delta_{ij} S_i$, we still find that the cross-spectra carry additional information with respect with the auto-spectra, in the sense:
\begin{equation}
    \label{eq:AutoCross}
    \hat{P}_{ii} \hat{P}_{jj} - \hat{P}_{ij}^{2} \to 
    \hat{P}_{ii}   S_j 
    + S_i \hat{P}_{jj}  
    +  S_i
    S_j
    \; .
\end{equation}
In this paper we show that, with multiple tracers, the accuracy with which we can measure the {\em irreducible} information in the cross-spectra, in the sense of Eq. \eqref{eq:AutoCross}, is not constrained by cosmic variance. 
Moreover, the uncertainties in those irreducible degrees of freedom of the cross-spectra fall even {\em faster} with the number density of tracers, when compared with the ratios of auto-spectra. Therefore, cosmic variance cancellation is not only a feature that allows us to measure bias, redshift-space distortion (RSD) parameters \cite{McDonald:2008sh} or primordial non-Gaussianities \cite{Seljak:2008xr} with increased accuracy, but it also opens the way to measure some of the parameters in the perturbative bias expansion to much higher accuracy by using multiple tracers and their cross-correlations.

We start, in Section 2, by deriving several useful results related to the multi-tracer Fisher matrix, with and without cross-spectra as independent degrees of freedom. 
We also present a general expression for the Fisher matrix with any data covariance, and show that its inverse is exactly the covariance matrix that we expect under the Gaussian approximation. 
Then, in Section 3 we show how to maximize the information of a galaxy survey, in terms of the optimal number tracers, by using two different summary statistics for the total amount of Fisher information in that survey. 
Finally, in Section 4 we show that the extra degrees of freedom that arise from the cross-spectra are not constrained by cosmic variance, and can be measured with arbitrary accuracy (at least in principle).
We conclude in Section 5.

\section{Fisher matrix for two-point functions}

Let's say that we have many samples of some measurements $f_i$, from which we wish to estimate physical parameters through the quadratic form (the ``correlations'') $q_{ij} \to \langle f_i f_j \rangle$. We define the data covariance as:
\begin{equation}
    \label{Eq:DataCov}
    C_{ij} = \langle f_i f_j \rangle = q_{ij} + s_{ij}\; ,
\end{equation}
where $s_{ij}$ is the noise (assumed symmetric under $i \leftrightarrow j$).
Under the hypothesis of Gaussianity for the data $f_i$, one can easily show by using Wick's theorem that the parameter covariance is given by:
\begin{equation}
    \label{Eq:DCov}
    {\rm Cov}(q_{ij}, q_{i'j'}) = 
    C_{ii'} C_{jj'} + C_{ij'} C_{ji'} \; .
\end{equation}

Since by construction $q_{ij} = q_{ji}$, this set of parameters will count twice the correlations. For this reason, we introduce the following notation for all the non-equivalent pairs $\{i,j\}$:
\begin{equation}
    \label{Eq:DoubleCount}
    q_{[ij]} = 
    q_{ij} \quad , \; {\rm for} \; i \leq j  \; .
\end{equation}
Clearly, if $i=1,2,\ldots,N$, then the  number of non-equivalent pairs is $N_p = N(N+1)/2$.
Notice that with this notation the parameter covariance is still given by the same expression as above, i.e.:
\begin{equation}
    \label{Eq:DCov2}
    {\rm Cov}(q_{[ij]}, q_{[i'j']}) = 
    C_{ii'} C_{jj'} + C_{ij'} C_{ji'} \; .
\end{equation}
However, as opposed to the $((N,N) \times (N,N))$ array of Eq. \eqref{Eq:DCov}, in terms of the individual pairs $[ij]$ the expression above is an $N_p \times N_p$ matrix.

With this notation it is straighforward to show that the inverse of the parameter covariance, also known as the Fisher matrix, is given by:
\begin{equation}
\label{eqn:FishMatGen}
    F[q_{[ij]},q_{[i'j']}] =
    F_{[ij],[i'j']} = 
    \left( 1 - \frac12 \delta^{ij} \right) 
    \left( 1 - \frac12 \delta^{i'j'} \right) \,
    \left( C_{ii'}^{-1} C_{jj'}^{-1} + C_{ij'}^{-1} C_{ji'}^{-1}  \right)
    \, .
\end{equation}
where $C_{ij}^{-1}$ is the inverse of the data covariance, i.e., $\sum_j C_{ij}^{-1} C_{ji'} = \delta_{ii'} $.

In order to show that the covariance of Eq. \eqref{Eq:DCov} is indeed the inverse of the Fisher matrix of Eq. \eqref{eqn:FishMatGen} we need the identity:
\begin{eqnarray}
\nonumber
    2 \sum_{[mn]} 
    C_{im}^{-1} C_{mi'} \, C_{jn}^{-1} C_{nj'} 
    &=& 
    \sum_{mn} 
    C_{im}^{-1} C_{mi'} \, C_{jn}^{-1} C_{nj'}
    +
    \sum_{m} 
    C_{im}^{-1} C_{mi'} \, C_{jm}^{-1} C_{mj'}
    \\ \nonumber
    &=& 
    \delta_{ii'} \, \delta_{jj'}
    +
    \sum_{m} 
    C_{im}^{-1} C_{mi'} \, C_{jm}^{-1} C_{mj'}
    \; ,
    \label{eqn:usefulid}
\end{eqnarray}
where, following the notation introduced in Eq. \eqref{Eq:DoubleCount}, the sum $\sum_{[mn]}$ is limited to the indices $m \leq n$.
With this result it is then trivial to show that, as expected:
\begin{equation}
    \label{Eq:FishCov}
    \sum_{[mn]} F_{[ij],[mn]} \, {\rm Cov}_{[mn],[i'j']}
    = \delta_{[ij],[i'j']} \, .
\end{equation}

We now present two key results for the covariance and Fisher matrices of the 2-point functions of Gaussian variables, which we will employ later. 
The first identity concerns the determinants of the Fisher matrix and of the parameter covariance (which are, of course, the inverse of each other). 
It is possible to show that:
\begin{equation}
    \det \left( {\rm Cov}_{[ij],[i'j']} \right) = 
    2^{N} \left( \det C \right)^{N+1} \; ,
\end{equation}
and therefore that:
\begin{equation}
    \label{Eq:detF}
    \det  \left( F_{[ij],[i'j']} \right) = 
    2^{-N} \left( \det C \right)^{-N-1} = 
    2^{-N} \left( \det C^{-1} \right)^{N+1} \; .
\end{equation}
We stress the fact that $F_{[ij],[i'j']}$ is an $N_p \times N_p$ matrix, while $C$ is an $N\times N$ matrix.

The second identity is the fact that the ``grand sum'' of the Fisher matrix is proportional to the square of the grand sum of the inverse data covariance, namely:
\begin{equation}
    \label{eq:TrF}
    \sum_{[ij]}  \sum_{[i'j']}   F_{[ij],[i'j']}  = 
    \frac12 \left( \sum_{ij} C^{-1}_{ij} \right)^2  \; .
\end{equation}
We will use Eqs. \eqref{Eq:detF} and \eqref{eq:TrF} in Section 3, when we optimize the number of tracers in a survey.

\subsection{Fisher matrix for the power spectra}
\label{sec:Fisher}

The fundamental degrees of freedom in a survey are the positions of the tracers (galaxies, halos or other point-like objects that follow the underlying matter distribution). 
When we measure the number densities of tracer species $i$, $n_i(\vec{x})$, over some volume around the position $\vec{x}$, that number reflects some mean density of those tracers, $\bar{n}_i(\vec{x})$, as well as the fluctuations $\delta n_i = n_i - \bar{n}_i$.
From these observables we compute the main object that carries information about cosmology, the data (or ``pixel'') covariance:
\begin{equation}
    \label{Eq:DefCorrFun}
    C_{ij} (\vec{x},\vec{y}) = \langle \delta n_i (\vec{x}) \, \delta n_j (\vec{y}) \rangle 
    = \bar{n}_i (\vec{x}) \,  \bar{n}_j (\vec{y}) \, \xi_{ij}(\vec{x},\vec{y}) + \bar{n}_i (\vec{x}) \delta_{ij} \delta_D (\vec{x} - \vec{y} ) \; ,
\end{equation}
where $\xi_{ij}(\vec{x},\vec{y}) $ is the 2-point correlation function, and the last term is shot noise, which we assume here to follow Poisson statistics.
The multi-tracer 2-point correlation function is generally assumed to be related to the matter correlation function, $\xi^{(m)}(\vec{x},\vec{y})$, through some knowable relations such as tracer bias, redshift-space distortions, etc. In real space (i.e., excluding redshift-space distortion), the matter two-point correlation function can be written in terms of the matter power spectrum as:
\begin{equation}
    \xi^{(m)}(\vec{x},\vec{y}) =
    \xi^{(m)}(|\vec{x} - \vec{y}|) =
    \int \frac{d^3 k}{(2\pi)^3} \, e^{-i \vec{k} \cdot (\vec{x}-\vec{y})} \, P^{(m)}(k) \; .
\end{equation}

We can also work directly in Fourier space, and derive the Fisher and covariance matrices for the power spectra. 
In that case it is more convenient to work with the density contrasts for the tracers, $\delta_i = (n_i-\bar{n}_i)/\bar{n}_i$.
The Fourier mode $\vec{k}$ of the density contrast can be expressed as:
\begin{equation}
\label{eqn:dof}
    d_i^a (\vec{k}) = \{ \tilde\delta_i (\vec{k}) \, , \, \tilde\delta_{i}^{*} (\vec{k}) \} \; ,
\end{equation}
where $i=1,2,\ldots,N$ denotes the tracer, and $a=1,2$ stand for the Fourier mode and its complex conjugate, respectively.
The data covariance is then:
\begin{equation}
\label{eqn:expval}
    \langle d_i^a (\vec{k}) d_j^b (\vec{k}{}') \rangle =  D^{ab}  \, C_{ij} (\vec{k},\vec{k}{}') = C^{ab}_{ij} (\vec{k},\vec{k}{}')\; ,
\end{equation}
where $D^{ab} = 1-\delta^{ab}$. 
The data covariance in Fourier space is then simply the observed power spectrum for those tracers, including shot noise if it is an auto-spectrum:
\begin{equation}
\label{eqn:expval2}
C_{ij} (\vec{k},\vec{k}{}') 
=  \delta_{\vec{k} \, \vec{k}{}'}  \left(  P_{ij} + \frac{\delta_{ij}}{ \bar{n}_i} \right) \; ,
\end{equation}
where in the continuum limit we have $\delta_{\vec{k} \vec{k}{}'} \to (2\pi)^3 \delta_D (\vec{k} - \vec{k}{}')$, but for simplicity here we can consider this to be a Kronecker delta-function, up to a constant.

One of the ways in which we can derive the Fisher matrix is through the Hessian of the log-likelihood.
Given a set of parameters $\theta^\mu$, the Fisher matrix is given by the generalized trace \cite{1997ApJ...480...22T}:
\begin{equation}
\label{eqn:FishMat}
    F_{\mu\nu} = \frac{1}{4} \sum_k V \tilde{V}_k \sum_{iji'j'} \sum_{aba'b'} 
    \frac{\partial \, C^{ab}_{ij}}{\partial \theta^\mu}
    \left[ C^{ba'}_{ji'} \right]^{-1} 
    \frac{\partial \, C^{a'b'}_{i'j'}}{\partial \theta^\nu} 
    \left[ C^{b'a}_{j'i} \right]^{-1} \; ,
\end{equation}
where $V$ is the survey volume, $\tilde{V}_k$ is the volume in Fourier space of the bandpowers (Fourier bins) $k$, and the additional factor of $1/2$ in Eq. (\ref{eqn:FishMat}) is due to the fact that our degrees of freedom include the Fourier modes twice.
Notice that this Fisher matrix is diagonal in $\vec{k}$ due to the diagonal nature of the data covariance, Eq. \eqref{eqn:expval2}, and for simplicity for the remainder of this Section we will omit the Fourier space indices.

Using the fact that the data covariance is separable,
$\left[ C^{ab}_{ij}\right]^{-1} = \left[ D^{ab}\right]^{-1}  C_{ij}^{-1}$, and using that $\left[ D^{ab}\right]^{-1} = D^{ab}$, we obtain $\sum_{abcd} D^{ab} D^{bc} D^{cd} D^{da} = \sum_{ac} \delta_{ac} \delta_{ca} = 2$. Hence:
\begin{equation}
\label{eqn:FishMat2}
    F_{\mu\nu} = \frac12 \sum_k 
    V \tilde{V}_k  \sum_{iji'j'}
    \frac{\partial \, C_{ij}}{\partial \theta^\mu}
    C_{ji'}^{-1} 
    \frac{\partial \, C_{i'j'}}{\partial \theta^\nu} 
    C_{j'i}^{-1} \; ,
\end{equation}
In Fourier space the inverse of the data covariance has a trivial expression, namely:
\begin{equation}
    \label{Eq:InvCovPk}
    C_{ij}^{-1} = 
    \bar{n}_i \, \delta_{ij} - 
    \bar{n}_i \, 
    \frac{P_{ij}}{1+{\cal{P}}} 
    \, \bar{n}_j \; ,
\end{equation}
with ${\cal{P}} = \sum_i \bar{n}_i \, P_{ii}$ -- in fact, the denominator in the second term is precisely $\det C = 1 + \cal{P}$.

We now set the parameters $\theta^\mu$ to be the auto- and cross-spectra of the tracers  evaluated at some bandpower, $P_{ij}(k)$.
Just as was the case in our basic example of the previous Section, these spectra are symmetric, $P_{ij}=P_{ji}$. In order to avoid double-counting these degrees of freedom we define the non-degenerate auto- and cross-spectra as:
\begin{equation}
    \label{Eq:sympow}
    P_{[ij]} = 
    P_{ij} \quad , \; {\rm for} \; i \leq j \; .
\end{equation}
Therefore, the spectra with any index can be expressed as:
\begin{equation}
    \label{Eq:ndpow}
    P_{ij} = P_{[ij]} + P_{[ji]} - \delta_{ij} P_{ii} \; ,
\end{equation}
and a similar expression for the data covariance.
We can now evaluate the partial derivatives assuming that the parameters are the non-denegerate spectra:
\begin{equation}
\label{Eq:InvCovk}
    \frac{\partial C_{ij}(k)}{\partial P_{[i'j']}(k')} = \delta_{k,k'} \left[ \delta_{ii'}\delta_{jj'}
+ \delta_{ij'}\delta_{ji'} - \delta_{ij}\delta_{ji'}\delta_{i'j'}\delta_{j'i} \right]
\, \equiv \,  \delta_{k,k'} \; \delta_{[ij],[i'j']}\; .
\end{equation}
Substituting this identity into Eq. (\ref{eqn:FishMat2}) results in a Fisher matrix which is diagonal in the bandpowers, and which can be expressed for each mode $k$ as:
\begin{equation}
\label{eqn:FishMat3}
    F[P^{[ij]},P^{[i'j']}] =
    F^{[ij],[i'j']} = V \tilde{V}_k  \,
    \left( 1 - \frac12 \delta^{ij} \right) 
    \left( 1 - \frac12 \delta^{i'j'} \right) \,
    \left( C_{ii'}^{-1} C_{jj'}^{-1} 
    + C_{ij'}^{-1} C_{ji'}^{-1}  \right)
    \, .
\end{equation}
It is immediately obvious that the results of the previous section apply here, so the inverse of this Fisher matrix is the usual expression for the covariance of the spectra:
\begin{equation}
    \label{eq:speccov}
    {\rm Cov} [P_{[ij]},P_{[i'j']}] =
    {\rm Cov}_{[ij],[i'j']} =
     \frac{1}{V \tilde{V}_k }\,
    \left( C_{ii'} C_{jj'} + C_{ij'} C_{ji'}  \right)
    \, .
\end{equation}

\subsection{Example: two tracers}

As an example, we write explicit expressions for the case when we have two tracers. The covariance matrix becomes:
\begin{equation}
    \label{eq:CovPk}
    {\rm Cov} [P_{[ij]},P_{[i'j']}] =
     \frac{1}{V \tilde{V}_k } \,
     \left(
     \begin{array}{ccccc}
     2 C_{11}^2 &  \quad & 2 C_{11} C_{12} &\quad & 2 C_{12}^2 \\
     {} & {} & {} & {} & \\
     2 C_{11} C_{12} &\quad & C_{11} C_{22} + C_{12}^2 &\quad & 2 C_{12} C_{22} \\
     {} & {} & {} & {} & \\
     2 C_{12}^2 &\quad & 2 C_{12} C_{22} &\quad & 2 C_{22}^2 
     \end{array}
    \right)
    \, ,
\end{equation}
where we have ordered the degrees of freedom as $\{ P_{11}, P_{12},P_{22} \} $.
The Fisher matrix in that case, derived directly from Eq. \eqref{eqn:FishMat3}, is given by:
\begin{eqnarray}
    \label{eq:FishPk}
    F [P_{[ij]},P_{[i'j']}] &=&
     V \tilde{V}_k  \,
     \left(
     \begin{array}{ccccc}
     \frac12 C_{11}^{-2} &  \quad & C_{11}^{-1} C_{12}^{-1} &  \quad &  \frac12 C_{12}^{-2} \\
     {} & {} & {} & {} & \\
     C_{11}^{-1} C_{12}^{-1} &  \quad &   C_{11}^{-1} C_{22}^{-1} + C_{12}^{-2}  &  \quad & C_{12}^{-1} C_{22}^{-1} \\
     {} & {} & {} & {} & \\
     \frac12 C_{12}^{-2} &  \quad & C_{12}^{-1} C_{22}^{-1} &  \quad & \frac12 C_{22}^{-2} 
     \end{array}
    \right) 
    \, .
\end{eqnarray}
In the particular case of the power spectra, the inverse of the data covariance is given by Eq. \eqref{Eq:InvCovPk}.

\section{Fisher information and the optimal number of tracers}

The results above can serve as a guide to a first attempt at organizing the data of galaxy surveys: is it worth splitting a galaxy population into sub-types with different properties, or upon doing so we risk degrading the discriminating power of our survey?
Although the precise answer depends on the nature of the tracers, as well as the kinds of parameters we are trying to constrain (in that respect see also \cite{2010MNRAS.407..772G}), there are some general trends that can be inferred in terms of the summary statistics of the Fisher matrix.

Before we proceed any further, it is useful to express the spectra in terms of signal-to-noise ratios (SNR), where the noise is given by shot noise (assumed Poissonian):
\begin{equation}
    \label{eq:defCalP}
    {\cal{P}}_{ij} = \sqrt{\bar{n}_i \, \bar{n}_j} \, P_{ij} \; .
\end{equation}
We define the the total clustering (or total SNR) as $ {\cal{P}} = \sum_i {\cal{P}}_{ii} =  \sum_i \bar{n}_i \, b_i^2 \, P^{(m)}$, and this quantity can be regarded as being approximately constant for a given survey.
When two tracer species $i$ and $j$ are joined to form a composite tracer, their numbers are combined,  $n_{i+j}= n_i + n_j$, but the number densities and linear biases are constrained by the relation $\bar{n}_{i+j} b_{i+j} = \bar{n}_{i} b_{i} + \bar{n}_{j} b_{j} $. 
Since ${\cal{P}} \sim \sum_i \bar{n}_i b_i^2$,  by joining or splitting tracers we can slightly increase or decrease the total SNR.

The Fisher and covariance matrices can also be expressed in terms of signal and noise. For generic degrees of freedom $X_\mu$ ($\mu=1,2,\ldots,n$) we have:
\begin{eqnarray}
    \label{eq:defCalF}
    {\cal{F}}[X_\mu,X_\nu] &=& X_\mu \, F [X_\mu,X_\nu] \, X_\nu =
    F[\log X_\mu , \log X_\nu ] \\
    \label{eq:defCalC}
    {\cal{C}}[X_\mu,X_\nu] &=& \frac{{\rm Cov}[X_\mu,X_\nu]}{X_\mu X_\nu} = {\rm Cov}[\log X_\mu, \log X_\nu] \; .
\end{eqnarray}
Clearly, these Fisher and covariance matrices refer to the relative uncertainties in the parameters $X_\mu$.
All our summary statistics will be derived in terms of these relative uncertainties -- or, equivalently, in terms of SNR.

There are arbitrarily many summary statistics that one can build from the Fisher matrix, and there is no single expression that can claim to capture the total information \cite{Bayes}. 
Although in principle we should be guided by invariants such as the trace or the determinant of the matrix, depending on the application one summary statistics may be more suitable.

The determinant of the Fisher matrix (whose square root is known as the Jeffreys prior \cite{Jeffrey}) is evidently a convenient summary statistics, and in the context of cosmology, when the Fisher matrix is projected into some sub-space, it is also called ``figure of merit'' (FoM) \cite{2006astro.ph..9591A}. In our context we define the FoM as:
\begin{equation}
    \label{eq:FoM}
    \Delta = \det  {\cal{F}} [X_\mu,X_\nu]  \; .
\end{equation}
The FoM corresponds to the inverse volume of an ellipsoid in $n$ dimensions, defined by the 68\% confidence limit of the corresponding multivariate Gaussian distribution. The smaller the volume in parameter space, the greater is the discriminating power.

Another useful summary statistics is a generalization of the $\chi^2$, given by the grand sum of the Fisher matrix. Let's take the usual $\chi^2$:
\begin{equation}
    \label{eq:chi2}
    \chi^2 = \sum_{\mu\nu} \left( X_\mu - \bar{X}_\mu \right) \, F[X_\mu,X_\nu] \, \left( X_\nu - \bar{X}_\nu \right) 
    = \sum_{\mu\nu} \left( 1 - \frac{\bar{X}_\mu}{X_\mu} \right) \,  {\cal{F}} [X_\mu,X_\nu] \, \left( 1 - \frac{\bar{X}_\nu}{X_\nu} \right)
    \; ,
\end{equation}
where $\bar{X}_\mu$ are the fiducial values of the parameters $X_\mu$. Now, make all $X_\mu$ equal to a certain fraction of the $\bar{X}_\mu$, such that the terms multiplying ${\cal{F}}$ in the sum above reduce to a constant. We then define the grand sum of the Fisher matrix as:
\begin{equation}
    \label{eq:defTotTrace}
    \Xi  
    = \sum_{\mu\nu} \, {\cal{F}} [X_\mu,X_\nu] 
    = \sum_{\mu\nu} \, X_\mu \, {\cal{F}} [X_\mu,X_\nu] \, X_\nu
    \; .
\end{equation}

%A related summary statistics follows the same basic argument, but now we use instead an anlogy with the $\chi^2$ per degree of freedom, and define: 
%\begin{equation}
%    \label{eq:defSigmadof}
%    \Sigma_{dof}  
%    = \frac{1}{n} \sum_{\mu\nu}
%    \, {\cal{F}} [X_\mu,X_\nu] 
%    \; .
%\end{equation}

\subsection{Fisher summary statistics for the auto-spectra}

Before we examine the Fisher matrix in full generality, it is instructive to consider the case when the cross-spectra are {\em not} independent degrees of freedom, but are in fact constrained in terms of the auto-spectra, $P_{ij}^2 = P_{ii} P_{jj}$. 
This corresponds to assuming that the non-linear and stochastic terms $\epsilon_i^S \to 0$ in Eq. \eqref{eq:deltai}, and that shot noise can be perfectly subtracted.
In that case the Fisher matrix is given by \cite{abramo2012full,abramo2013multitracer}:
\begin{equation}
    \label{eq:FishMT}
    {\cal{F}}[{\cal{P}}_{ii},{\cal{P}}_{jj}]
    = \frac{V \tilde{V}_k}{4}
    \frac{\delta_{ij} {\cal{P}}_{ii} {\cal{P}} (1+{\cal{P}}) + {\cal{P}}_{ii} {\cal{P}}_{jj} (1-{\cal{P}})}{(1+{\cal{P}})^2} \; .
\end{equation}
It is worth stressing here that by including only the auto-spectra in the Fisher matrix does not mean that we are discarding measurements of the cross-spectra: it simply means that the degrees of freedom in the cross-spectra are assumed to be redundant with those already found in the cross-spectra.

It is straightforward to show that the FoM associated with the Fisher matrix of Eq. \eqref{eq:FishMT} is given by:
\begin{equation}
    \label{eq:detFauto}
    \Delta = 
    \left( \frac{V \tilde{V}_k}{4}  \right)^N 
    \,
    \frac{2 \, {\cal{P}}^N }{(1+{\cal{P}})^{N+1}}
    \, \prod_i {\cal{P}}_{ii}  \; .
\end{equation}
We can now ask what happens as we change the number of tracers -- either by splitting one tracer into two or more sub-types, or by combining many tracers into a single one. 
However, notice that the determinant of Eq. \eqref{eq:detFauto} carries a phase-space volume factor for each degree of freedom, $(V\tilde{V}_k)^N$.
Moreover, when projecting into a final set of parameters, those phase space volumes will tend to be compensated by the larger number of parameters from each tracer species.
This discussion indicates that what we ought to use as a summary statistics is the Fisher information per unit phase space volume, which in this case (where we ignore the information from the cross-spectra) results in the renormalized FoM:
\begin{equation}
    \label{eq:detFautoRenorm}
    \Delta_{Ph} = \frac{\Delta_{N}}{ \left( V \tilde{V}_k\right)^N}
    =
    \frac{1}{4^N}
    \,
    \frac{2 \,  {\cal{P}}^N}{(1+{\cal{P}})^{N+1}}
    \, \prod_i {\cal{P}}_{ii}  \; .
\end{equation}

In order to maximize the Fisher information it is clear that we must first maximize the clustering SNR ${\cal{P}}$, given the constraint that the total number of tracers is kept fixed -- this is indeed what we expect from a survey where we observe some total number of objects, that we can  subdivide into one or more species of tracers. But for a fixed number of tracers ($N$), what is the optimal way to draw the lines between the tracers? 

The answer clearly depends on the way that bias varies with the number density.
Let's assume that the bias is given by a power-law in terms of the number density: $b_i^2 \sim \bar{n}_i^{-\gamma}$. 
The total SNR is therefore given by:
\begin{equation}
    \label{Eq:Pofgamma}
    {\cal{P}}  = \sum_i {\cal{P}}_{ii} 
    = \sum_i \bar{n}_i \, b_i^2 \, P^{(m)} 
    = \bar{n}_T \, b_T^2 \, P^{(m)}  \sum_i w_i^{1-\gamma} \; ,
\end{equation}
where $\bar{n}_T = \sum_i \bar{n}_i$,
$b_i^2 = b_T^2 (\bar{n}_i/\bar{n}_T)^{-\gamma}$,
and we defined the weights:
\begin{equation}
    \label{Eq:weights}
    w_i = \frac{\bar{n}_i}{\bar{n}_T} \; .
\end{equation}

Extremizing the total clustering SNR subject to the constraint that the total number of objects is fixed ($\sum_i w_i = 1$) leads to an ``equipartition'' between all the tracers, i.e., $w_i = 1/N$.
This extremal value is a maximum only if $\gamma (1-\gamma) < 0$, or $0 < \gamma < 1$ -- otherwise equipartition yields a minimum.
Substituting $w_i=1/N$ back into Eq. \eqref{Eq:Pofgamma} we obtain that 
${\cal{P}} \to \bar{n}_T b_T^2 P^{(m)}  N^\gamma$, which grows with the number of tracers if the power-law $\gamma$ is in the range where equipartition yields a maximum of the clustering SNR.

Interestingly, the halo mass function and halo bias fits found by Tinker et al. \cite{2008ApJ...688..709T} indicate that, for the halos masses of cosmological interest ($10^{10} \lesssim M[h^{-1} M_\odot] \lesssim 10^{15}$), this index is $0.1 \lesssim \gamma \lesssim 0.5 $. 
Therefore, for tracers which behave  similarly to dark matter halos, the configuration which maximizes the total SNR ${\cal{P}}$ has the populations divided in equal numbers, with as many sub-species of tracers as practically possible.
If, on the other hand, the tracers behave in a completely different way, such that $\gamma <0$ or $\gamma > 1$, then the maximum of ${\cal{P}}$ is obtained by taking one of the $w_i \to 1$, and all others to zero -- i.e., in such a situation it would be optimal to join all the tracers into a single population. For mixed tracers, such as galaxies, separated by  properties such as stellar mass, luminosity, morphology, etc., as long as the parameter used to separate the populations generates a dependence of the bias with number density which follows the same trends as the dark matter halos, then the main conclusion remains the same.

Notice that equipartition in terms of the number densities, which maximizes ${\cal{P}}$, also implies equipartition of that total clustering SNR amongst all tracers: since $w_i = 1/N$, we have that $w_i^{1-\gamma} = N^{\gamma-1}$, and therefore ${\cal{P}}_{ii} = {\cal{P}}/N$.
Moreover, equipartition of the clustering also happens to maximize the product $\prod_i {\cal{P}}_{ii}$, which implies that the FoM  of Eq. \eqref{eq:detFautoRenorm} is maximized when $w_i = 1/N$, and ${\cal{P}}_{ii} = {\cal{P}}/N$.

It is more enlightening to express the FoM at its maximum value in terms of the total SNR ${\cal{P}}$.
Substituting the optimal configuration (i.e., equipartition) into Eq. \eqref{eq:detFauto} we obtain:
\begin{equation}
    \label{eq:MaxdetFauto}
    \Delta_{Ph}^{\rm max} =  
    \frac{1}{4^N}
    \frac{2 \,  {\cal{P}}^N}{(1+{\cal{P}})^{N+1}}
    \, \left( \frac{{\cal{P}}}{N} \right)^N \; .
\end{equation}
While in this expression the total SNR ${\cal{P}}$ still depends on the number of tracers through a relation such as ${\cal{P}} \sim N^\gamma$, we can find an approximate expression for the number of tracers that maximize the FoM, assuming that ${\cal{P}}$ varies slowly with $N$ -- i.e., in the limit that $\gamma$ is small. We have, in that limit:
\begin{equation}
    \label{eq:NmaxDet}
    \left. \frac{d \Delta_{Ph}^{\rm max}}{dN} \right|_{{\cal{P}}} = 0 
    \quad \Rightarrow \quad
    N_0^{\rm max} 
    = \frac{1}{4\, e}  \, 
    \frac{{\cal{P}}^2}{1+{\cal{P}}} \; ,
\end{equation}
where $e$ is Euler's number.
Clearly, then, we obtain that, as the total clustering strength ${\cal{P}}$ increases, so grows the optimal number of tracers that we should use in order to maximize the information from our survey. 

\begin{figure}
    \centering
    \includegraphics[width=0.48\textwidth]{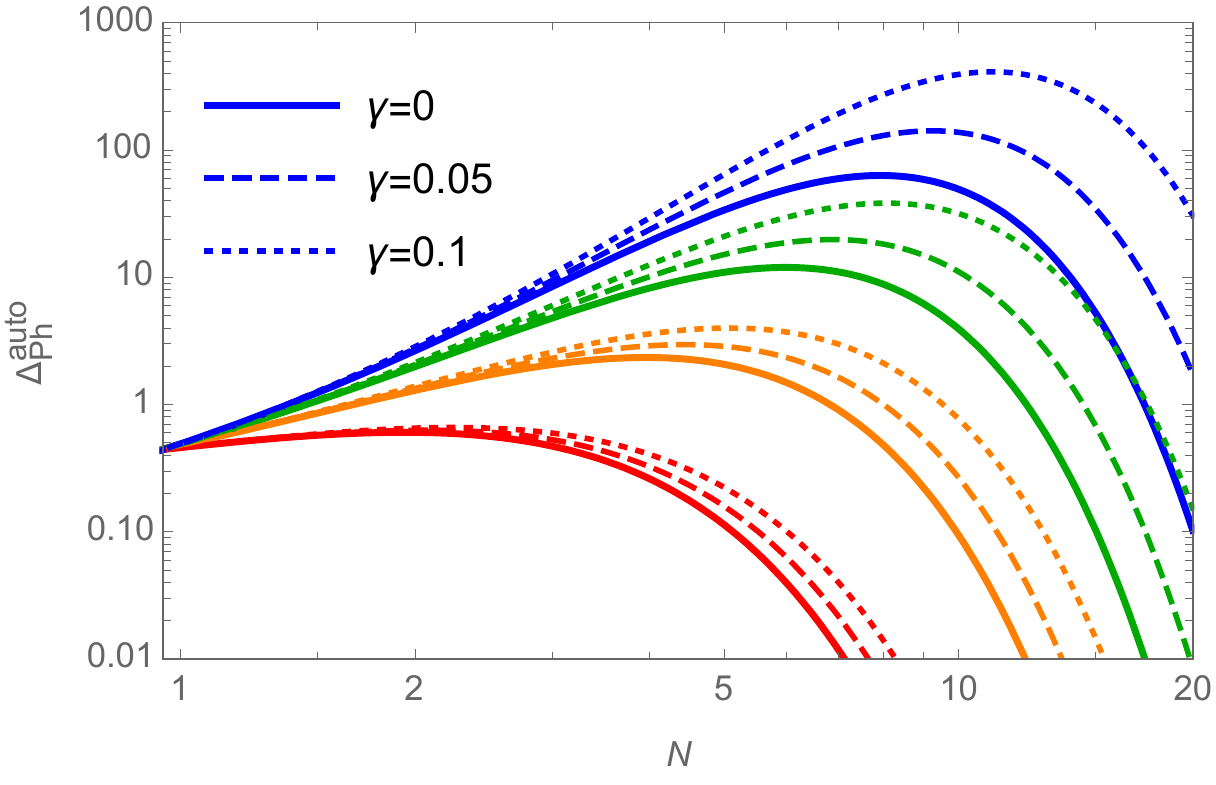}
    \includegraphics[width=0.48\textwidth]{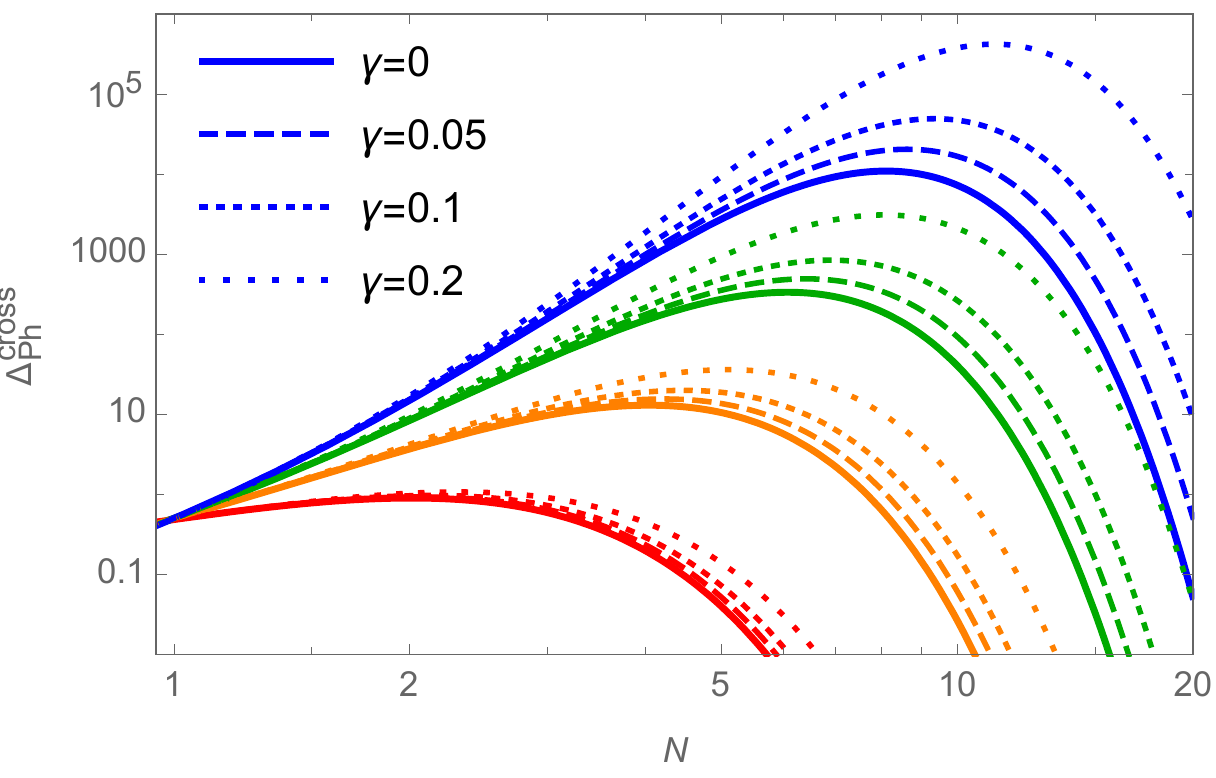}
    \caption{Figure of Merit (FoM) as a function of the number of tracers, $N$, assuming that ${\cal{P}} = P_0 N^\gamma$. 
    Left panel: FoM considering only the auto-spectra, Eq. \eqref{eq:detFautoRenorm}, for the power law indices $\gamma=0$ (solid), 0.05 (dashed), and 0.1 (thin-dashed lines).
    The different colors refer to different values of $P_0$ that make $N_0^{max}=2$ (red), 4 (orange), 6 (green) and 8 (blue lines) -- see Eq. \eqref{eq:NmaxDet}.
    Right panel: FoM considering the auto- and cross-spectra, Eq.  \eqref{Eq:DetFCGen}, for the power law indices $\gamma=0$, 0.05, 0.1 and 0.2, which are respectively denoted by the solid, dashed, thin-dashed and dotted lines. As in the left panel, the red, orange, green and blue lines refer to the values of $P_0$ that make $N_0^{max}=2$, 4, 6 and 8 -- in this instance, see Eq. \eqref{eq:NMaxDet2}.}
    \label{fig:Deltas}
\end{figure}

In a more realistic situation, we cannot hold the total clustering SNR ${\cal{P}}$ constant as we change the number of species of tracers. 
In Fig. \ref{fig:Deltas} (left panel) we show the FoM as a function of $N$, assuming that ${\cal{P}} = P_0 N^\gamma$ --- here $P_0=\bar{n}_T b_T^2 P^{(m)} $ is the baseline clustering SNR\footnote{E.g., a low-redshift survey with $\bar{n}_T = 2. \times 10^{-3} \, h^3$ Mpc$^{-3}$ and a mean bias $b_T =1.5$, at a reference scale of $k_0=0.1 \, h$ Mpc$^{-1}$, and considering a spectrum $P^{(m)} (k_0)=10^4 \, h^{-3}$ Mpc$^3$, yields $P_0 =45$.}.
The solid, dashed and thin-dashed lines denote the power-law values $\gamma=0$, 0.05 and 0.1, while the colors refer to the values of $P_0$ which make $N^0_{max}=$ 2 (red, $P_0=22.2$), 4 (orange, $P_0=44$), 6 (green, $P_0=65.7$) and 8 (blue, $P_0=87.5$), according to Eq.  \eqref{eq:NmaxDet} -- i.e., assuming that $\gamma \to 0$.
This plot shows that the higher the clustering SNR $\cal{P}$ is, the higher is also the optimal value for the number of tracers, $N^{max}$.
The plot also shows that, for a fixed value of the baseline clustering $P_0$, the optimal number of tracers grows slightly for higher values of $\gamma$ (as denoted by the shift to higher values of $N$ of the dashed lines with respect to the solid lines).

As an alternative to the determinant of the Fisher matrix, we could also consider the grand sum defined in Eq. \eqref{eq:defTotTrace} as a summary statistics for the total Fisher information. 
In the case where the auto-spectra contain all the information, that grand sum is obtained directly from Eq. \eqref{eq:FishMT}:
\begin{equation}
    \label{eq:trFauto}
    \Xi =  
    \frac{V \tilde{V}_k}{2}
    \,
    \frac{{\cal{P}}^{2}}{(1+{\cal{P}})^{2}} \; .
\end{equation}
Interestingly, this summary statistics does not carry an explicit dependency on the number of tracers: for a fixed volume and bandwith, the total information depends only on the total clustering SNR ${\cal{P}}$.
Since maximizing that SNR also maximizes the grand sum, this leads us back to the same conclusion that equipartition is the optimal solution.

%Therefore, it seems more intuitive to consider the grand sum divided by the number of degrees of freedom as per Eq. \eqref{eq:defSigmadof}, which in this case leads to:
%\begin{equation}
%    \label{eq:SigmaFauto}
%    \Sigma_{N,dof} =  
%    \frac{V \tilde{V}_k}{2 N}
%    \,
%    \frac{{\cal{P}}^{2}}{(1+{\cal{P}})^{2}} \; .
%\end{equation}
%Therefore, from the standpoint of this summary statistics the Fisher information per degree of freedom scales with the inverse of the number of tracers, while the total information is fixed.

\subsection{Fisher summary statistics: including the cross-spectra}

We now include the independent information from the cross-spectra in our analysis.
In that case, the determinant of the Fisher matrix can be derived directly from Eq. \eqref{Eq:detF}. 

Here we assume, {\em a posteriori}, that for the fiducial model we have ${\cal{P}}_{12}^2 \to {\cal{P}}_{11} {\cal{P}}_{22}$. This means that in our expressions we assume that the non-linear and stochastic terms $\epsilon_i$ are relatively small -- a good approximation on large scales.
The resulting FoM is then given by:
\begin{eqnarray}
    \label{Eq:DetFCGen}
    \Delta =
    \det \left( {\cal{F}}_{[ij],[i'j']} \right) &=& 
    2^{-N} \left( V \tilde{V}_k  \right)^{N(N+1)/2} \frac{\prod_{i} {\cal{P}}_{ii}^2}{\left( 
    1+{\cal{P}} \right)^{N+1}}  \; . 
\end{eqnarray}
This expression should be compared with Eq. \eqref{eq:detFauto}, in the case when the cross-spectra are not independent degrees of freedom. 
Once again, there are multiplicative factors of the phase space volume for each one of the degrees of freedom -- in this case, the $N(N+1)/2$ auto- and cross-spectra. The renormalized FoM then becomes:
\begin{eqnarray}
    \label{Eq:DetFCGenNorm}
    \Delta_{Ph} =
    &=& 
    2^{-N} \frac{\prod_{i} {\cal{P}}_{ii}^2}{\left(
    1+{\cal{P}} \right)^{N+1}}  \; . 
\end{eqnarray}
Just as it happened in the previous Section, maximizing the FoM leads to the equipartition of power between the tracers, ${\cal{P}}_{ii}  \to {\cal{P}}/N$ -- and the same conditions still apply regarding the dependence of the biases of the tracers with the number densities, which we assume to be $b_i^2 \sim \bar{n}_i^{-\gamma}$, with $0<\gamma <1$ so that the extremum of the FoM is in fact a maximum.

Finally, knowing that the equipartition of power maximizes the FoM $\Delta_{N,Ph}$, we can ask whether or not it is beneficial to split a sample into sub-types of tracers. 
Let's again consider the FoM at its maximum, i.e., substituting $P_{ii} = {\cal{P}}/N$ into Eq. \eqref{Eq:DetFCGenNorm}:
\begin{eqnarray}
    \label{Eq:DetFCGenMax}
    \Delta_{Ph}^{\rm max} = \frac{1}{2^{N} \, N^{2N} }
    \frac{{\cal{P}}^{2N}}{ \left( 
    1+ {\cal{P}}\right)^{N+1}}  \; .
\end{eqnarray}

In the right panel of Fig. \ref{fig:Deltas} we show the FoM of Eq. \eqref{Eq:DetFCGenMax} as a function of $N$, again assuming that ${\cal{P}} = P_0 N^\gamma$. In this case we have plotted the FoM for the values $\gamma=0$ (solid), 0.05 (dashed), 0.1 (thin-dashed), and 0.2 (dotted lines), and the colors refer to the values of $P_0$ which make $N_0^{max}=$ 2 (red, $P_0=60$), 4 (orange, $P_0=237$), 6 (green, $P_0=533$) and 8 (blue, $P_0=947$), according to Eq.  \eqref{eq:NMaxDet2}.
As was the case when we eliminated the information in the cross-spectra (left panel of Fig. \ref{fig:Deltas},) we see that higher values of the clustering SNR $\cal{P}$ lead to higher values for the optimal number of tracer species.

However, there are some important differences when we include the cross-spectra. 
First, for similar values of the baseline SNR $P_0$, the optimal number of tracers when we disregard the degrees of freedom of the cross-spectra is higher than the optimal number of tracers when we take into account those degrees of freedom: e.g, compare the green lines in the left plot of Fig. \ref{fig:Deltas} (for $P_0 = 65.7$) with the red lines in the plot on the right (for $P_0 = 65.7$). 
Discarding the cross-spectra, we would arrive at an optimal number of $N\sim 8-10$ tracers if we neglected the cross-spectra, and an optimal number of $N\sim 2-3$ tracers if we fully incorporate those degrees of freedom.
Another way of putting this is to notice that, when including the cross-spectra, the baseline clustering SNR needs to be significantly higher for the same optimal number of tracers compared with the case where the cross-spectra are not independent degrees of freedom: e.g., for $N^{max}_0=4$ we only need $P_0=44$ when the cross-spectra are discarded, while that value grows to $P_0=237$ when they are included.

The second difference is that, when we include the cross-spectra, the FoM becomes less sensitive to the power-law index $\gamma$, as can be seen by the difference between the solid, dashed and dotted lines of the plot in the right panel of Fig. \ref{fig:Deltas}, which is much less pronounced than in the case when we discarded the degrees of freedom in the cross-spectra (left panel of the same figure).

\begin{figure}
    \centering
    \includegraphics[width=0.7\textwidth]{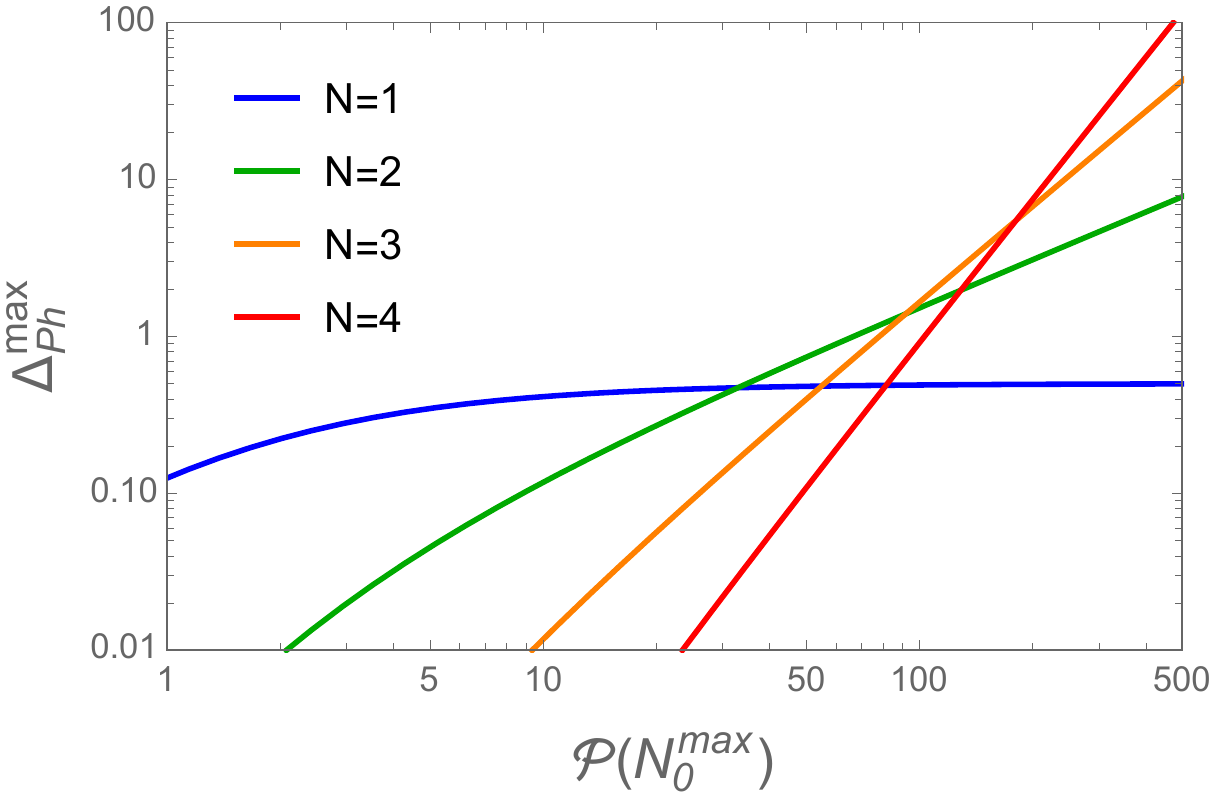}
    \caption{Figure of Merit (FoM) for the Fisher information at its maximum, $\Delta_{N,Ph}^{max}$, as a function of the total power ${\cal{P}}$, for 1 (blue), 2  (green), 3  (orange), and 4 (red) tracers.
    Here we assumed that ${\cal{P}}$ does not vary significantly with $N$ -- see the text.}
    \label{fig:totfish}
\end{figure}

Just as we did in the previous Section, an approximate expression for the optimal number of tracers can be found by taking the derivative of the FoM while assuming that the clustering SNR is kept fixed:
\begin{eqnarray}
    \label{eq:NMaxDet2}
    \left. \frac{d \, \Delta_{Ph}^{\rm max}}{dN} \right|_{{\cal{P}}} = 0 
    \quad \Rightarrow \quad 
    N_0^{max} = \frac{1}{\sqrt{2} \, e} \, 
    \frac{{\cal{P}}}{ 
    \sqrt{1+ {\cal{P}}}}  \; .
\end{eqnarray}
Since in the case where the cross-spectra are included the FoM is not very sensitive to $\gamma$ (see Fig. \ref{fig:Deltas}), we can take the limit $\gamma \to 0$ and substitute the maximal number found in Eq. \eqref{eq:NMaxDet2} into Eq. \eqref{Eq:DetFCGenMax}, to express the FoM as a function of the total SNR ${\cal{P}}$. 
The result is shown in
Fig. \ref{fig:totfish}, for $N=1$, 2, 3, and 4.

The other summary statistics that we can employ is the grand sum of the relative Fisher matrix, Eq. \eqref{eq:defTotTrace}. We have, for the relative Fisher matrix of the auto- and cross-spectra:
\begin{equation}
    \label{eq:TrF2}
    \Xi
    =
    \sum_{[ij]}  \sum_{[i'j']}   {\cal{F}}_{[ij],[i'j']}  = 
    \frac12 \, V \tilde{V}_k \, 
    \left( N - \frac{1}{1+{\cal{P}}} 
    \sum_{ij} {\cal{P}}_{ij} \right)^2  \; .
\end{equation}
We can again extremize this functional assuming that $b_i^2 \sim \bar{n}_i^{-\gamma}$ ,  subject to the constraint $\sum_i w_i = \sum_i \bar{n}_i/\bar{n}_T = 1$.
The result is the same as before: the weights are all the same, meaning equipartition of the number densities and SNRs, $w_i = 1/N$ and
${\cal{P}}_{ij} ={\cal{P}}_{ii} = {\cal{P}}/N$. Substituting this extremal solution back into Eq. \eqref{eq:TrF2} we obtain:
\begin{eqnarray}
    \label{Eq:DetFCGenMax2}
    \Xi^{\rm max} &=& \frac12 \,
    V \tilde{V}_k \, 
    \left( N - \frac{1}{1+{\cal{P}}} N^2 \frac{{\cal{P}}}{N}
    \right)^2 \\ \nonumber
    &=&   \frac12 \, V \tilde{V}_k \, 
    \frac{N^2}{\left( 1+{\cal{P}}\right)^2} \; .
\end{eqnarray}
Now, recall that ${\cal{P}} \sim N^\gamma$, with $0.1 \lesssim \gamma \lesssim 0.5$, and then it becomes clear that the grand sum Fisher information always grows when we split the tracers into more sub-types.

%However, again we would be better served by analyzing the information per degree of freedom, which is given by:
%\begin{eqnarray}
%    \label{Eq:DetFCGenMax3}
%    \Sigma_{N,dof}^{\rm max} 
%    &=&   \frac{V \tilde{V}_k}{N(N+1)} \, 
%    \frac{N^2}{\left( 1+{\cal{P}}\right)^2} \; .
%\end{eqnarray}
%Therefore, we are constrained by diminishing returns: increasing the number of tracers by one, $N\to N+1$, increases the Fisher information $\Sigma_{N,dof}^{\rm max}$ only by $(N+2)^{-1}(N+1)^{-1}$. 
%Of course, one may still achieve large gains in the Fisher information if the power is not optimally shared between the tracers.

A caveat is in order with regards to the conclusions that were drawn above.
The precise behaviours of the FoM and the grand sum Fisher information depend on the nature of the relationship between the number densities of the tracers, $\bar{n}_i$, and their biases, $b_i$. 
In the derivation above we assumed that $b_i^2 \sim \bar{n}_i^{-\gamma}$ for all tracers, with $0 < \gamma < 1$, regardless of how the tracers are split. 
But this is not strictly valid even for halos: at $z=0$ the fit by Tinker for halo bias at low ($ M_h  \lesssim 10^{12} \, h^{-1} \, M_\odot$) and high ($ M_h \gtrsim 10^{15} \, h^{-1} \, M_\odot$) masses yields $\gamma \sim 0.1-0.3$, while in the intermediate mass range $\gamma \sim 0.4-0.5$ \cite{2008ApJ...688..709T}.
For galaxies, quasars and Ly-$\alpha$ systems the way in which we can break up the tracers into sub-types, according to luminosity, stellar mass or other properties, could be even less consistent with the hypothesis of a simple power-law relationship between number densities and biases. If that turns out to be the case in a given galaxy survey, then maximizing the information may be significantly more complex than in this simplified model.

\section{The independent degrees of freedom: diagonalizing the Fisher matrix}

If the cross-correlations do not carry any additional information, then, as shown by \cite{abramo2013multitracer}, the independent degrees of freedom which diagonalize the Fisher matrix of Eq. \eqref{eq:FishMT} are given by the total clustering power, ${\cal{P}}$, and ratios of their clusterings.
In the case of two tracers, the diagonalized Fisher matrix for the degrees of freedom $\{ {\cal{P}} \, , \log ({\cal{P}}_{11} / {\cal{P}}_{22}) \} $ is given by:
\begin{eqnarray}
    \label{eq:FishDiag}
    F [{\cal{P}} \, , \, \log ({\cal{P}}_{11} / {\cal{P}}_{22})] &=&
     V \tilde{V}_k  \,
     \left(
     \begin{array}{cc}
     \frac12  \frac{1}{(1+{\cal{P}})^2} &  0  \\
     0 &  \frac14  \frac{{\cal{P}}_{11} \, {\cal{P}}_{22} }{1+{\cal{P}}} 
     \end{array}
    \right) 
    \, .
\end{eqnarray}
This expression makes it clear that, at least in the Gaussian approximation, measurements of the total clustering strength ${\cal{P}}$ are limited by cosmic variance: even if the total clustering becomes arbitrarily large, ${\cal{P}} \to \infty$, its relative uncertainty is still limited by the volume of the survey and the Fourier-space volume of the bandwith, $\sigma^2({\cal{P}})/{\cal{P}}^2 \to 2/(V \tilde{V}_k)$. The ratios of spectra, on the other hand, can be measured with arbitrarily large accuracy (at least in principle), as long as we keep increasing the number densities of the two tracers.

However, it is only on very large scales that one can realistically expect that the auto- and cross-spectra are degenerate,
$P_{ij}^2 = P_{ii} P_{jj}$. Indeed, as argued in the Introduction, on small scales the cross-spectra may carry information about additional physical dependencies that are not directly available through the auto-correlations.
But more importantly, cross-correlations and cross-spectra constitute different observables, which can be estimated from the data in different ways, in order to optimize the amount of information that is extracted from the survey. 

Therefore, the question arises as to what are the independent degrees of freedom when the cross-spectra are regarded as carrying irreducible degrees of freedom, which are not degenerate with the auto-spectra.
To be specific, the problem we wish to solve, in the particular case of two tracers, is how to diagonalize the Fisher matrix of Eq. \eqref{eq:FishPk}.
A straightforward calculation shows that the three degrees of freedom which diagonalize that Fisher matrix are:
\begin{eqnarray}
    \label{eq:Q1}
    {\cal{Q}}_1 &=&  
    {\cal{P}} \left[ 1 + \frac12 \log \left( 
    \frac{{\cal{P}}_{11}^2 + {\cal{P}}_{22}^2 + 2 {\cal{P}}_{12}^2}{{\cal{P}}^2} 
    \right) \right] \\ 
    \label{eq:Q2}
    {\cal{Q}}_2 &=& \log \left( 
    \frac{{\cal{P}}_{11}^2 + {\cal{P}}_{12}^2 }{{\cal{P}}_{22}^2 + {\cal{P}}_{12}^2} 
    \right) \\
    \label{eq:Q3}
    {\cal{Q}}_3 &=& 
    \frac{{\cal{P}}_{12}^2 - {\cal{P}}_{11} {\cal{P}}_{22}}{{\cal{P}}^2 } \; .
\end{eqnarray}
It is clear that, in the limit that ${\cal{P}}_{12}^2 \to {\cal{P}}_{11} {\cal{P}}_{22}$, the first two degrees of freedom reduce to ${\cal{Q}}_1 \to  {\cal{P}}$ and to ${\cal{Q}}_2 \to \log {\cal{P}}_{11}/{\cal{P}}_{22}$, while the third one effectively disappears, ${\cal{Q}}_3 \to 0$.
For that reason, it is useful to express the irreducible degree of freedom in the cross-spectrum in terms of an adimensional quantity $\epsilon_{12}$:
\begin{equation}
    {\cal{P}}_{12}^2 = {\cal{P}}_{11}  {\cal{P}}_{22} (1 + \epsilon_{12} ) 
    \quad \Leftrightarrow \quad
    {\cal{Q}}_3  = \frac{{\cal{P}}_{11}  {\cal{P}}_{22}}{{\cal{P}}^2} \, \epsilon_{12} 
    \, .
\end{equation}

Computing the Jacobian for the transformation from $\{ {\cal{P}}_{11} ,{\cal{P}}_{12} ,{\cal{P}}_{22} \} \to \{ {\cal{Q}}_1 , {\cal{Q}}_1 , {\cal{Q}}_3\} $ and using it to project the Fisher matrix of Eq. \eqref{eq:FishPk} into the new degrees of freedom, yields the result:
\begin{eqnarray}
    \label{eq:DiagFishPk}
    F [{\cal{Q}}_i , {\cal{Q}}_j]&=&
     V \tilde{V}_k  \,
     \left(
     \begin{array}{ccccc}
     \frac12  \frac{1}{(1+{\cal{P}})^2}  &  \quad & 0 &  \quad & 0 \\
     {} & {} & {} & {} & \\
     0 &  \quad &  \frac14  \frac{{\cal{P}}_{11} \, {\cal{P}}_{22} }{1+{\cal{P}}}  &  \quad & 0 \\
     {} & {} & {} & {} & \\
     0 &  \quad & 0 &  \quad & \frac12 (1+{\cal{P}})^2
     \end{array}
    \right) 
    \, .
\end{eqnarray}
Notice that our {\em fiducial model} is such that $P_{12}^2 \to P_{11} P_{22}$, or $\epsilon_{12} \to 0$, but we only take this limit {\em after} transforming to the new degrees of freedom.

\begin{figure}
    \centering
    \includegraphics[width=0.7\textwidth]{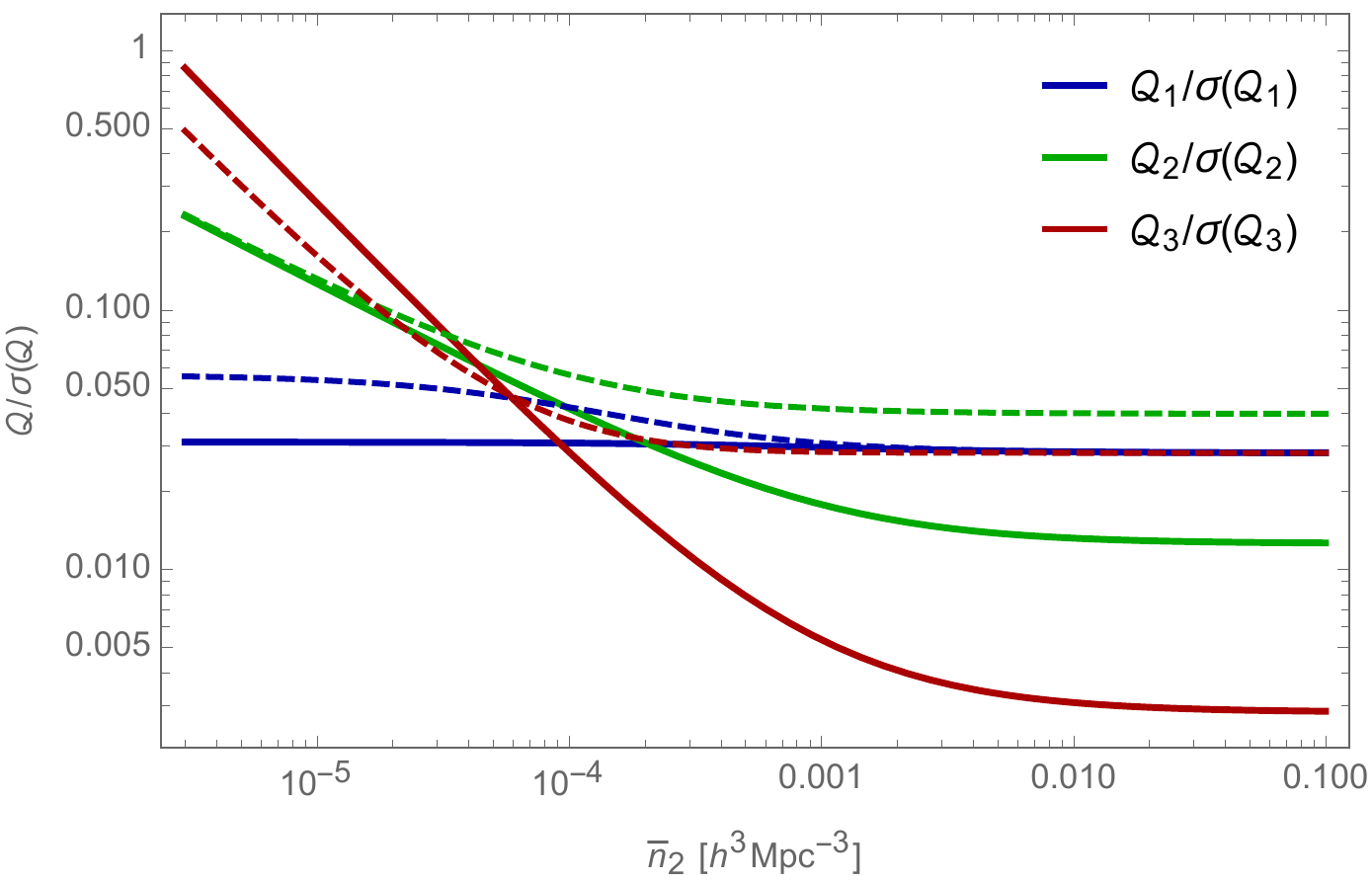}
    \caption{Signal-to-noise ratios ${\cal{Q}}_i/\sigma({\cal{Q}}_i)$, for the three independent degrees of freedom in the case of two tracers -- see Eqs. \eqref{eq:Q1}-\eqref{eq:Q3}.
    The blue, green and red curves correspond to the SNRs of ${\cal{Q}}_1$, ${\cal{Q}}_2$ and ${\cal{Q}}_3$. 
    The thin, dashed lines correspond to the scenario with $\bar{n}_1 = 10^{-4} \, h^{3} \, {\rm Mpc}^{-3}$, while the thick, solid lines correspond to the case where $\bar{n}_1 = 10^{-3} \, h^{3} \, {\rm Mpc}^{-3}$, both as a function of $\bar{n}_2$. 
    For these examples 
    we considered a volume $V=10^9 \, h^{-3} \, {\rm Mpc}^3$, a spherical bandpower at $k= (0.1 \pm 0.0025) \, h \, {\rm Mpc}^{-1}$, and a value of the power spectrum of $P^{(m)} (k) = 10^4 \, h^{-3} \, {\rm Mpc}^3$.}
    \label{fig:SNR}
\end{figure}

Some facts are immediately clear: first, the two generalized degrees of freedom ${\cal{Q}}_1$ and ${\cal{Q}}_2$, have the same Fisher information as the simplified degrees of freedom ${\cal{P}}$  and $\log ({\cal{P}}_{11}/{\cal{P}}_{22})$ -- see Eq. \eqref{eq:FishDiag}.
But more importantly, just as it happens with the relative clustering strength ${\cal{Q}}_2$, the irreducible information in the cross-spectrum, encapsulated in 
${\cal{Q}}_3$, also have a signal-to-noise ratio (SNR) that grows arbitrarily with the number density of the tracers, i.e.: 
\begin{equation}
\label{Eq:SigmaQ3}
    \frac{{\cal{Q}}_3^2}{\sigma^2({\cal{Q}}_3)} 
    = \frac{V \tilde{V}_k}{2}
    (1+{\cal{P}})^2 
    \left(
    \frac{{\cal{P}}_{12}^2 - {\cal{P}}_{11} {\cal{P}}_{22}}{{\cal{P}}^2} \right)^2
    = \frac{V \tilde{V}_k}{2}
    (1+{\cal{P}})^2 
    \left(
    \frac{ {\cal{P}}_{11} {\cal{P}}_{22}}{{\cal{P}}^2} \right)^2
    \epsilon_{12}^2 
    \, .
\end{equation}
Fig. \ref{fig:SNR} shows the behavior of the SNRs of the three independent degrees of freedom in some typical scenarios.

Taking the high SNR limit, ${\cal{P}}_1  = \bar{n}_1 P_{11} \gg 1$ and
${\cal{P}}_2 = \bar{n}_2 P_{22} \gg 1$, we see that the accuracies in the measurements of the three independent variables scale as:
\begin{eqnarray}
    \label{eq:SNR1}
    \frac{{\cal{Q}}_1^2}{\sigma^2({\cal{Q}}_1)} &\sim & 
    \left( \frac{\bar{n}_1 \, \bar{n}_2}{\bar{n}_1 + \bar{n}_2}  \right)^0
    \to \left( \frac{\bar{n}_T}{4} \right)^0
    \\ 
    \label{eq:SNR2}
    \frac{{\cal{Q}}_2^2}{\sigma^2({\cal{Q}}_2)} &\sim & 
    \left( \frac{\bar{n}_1 \, \bar{n}_2}{\bar{n}_1 + \bar{n}_2}  \right)^1
    \to \left( \frac{\bar{n}_T}{4} \right)^1
    \\     
    \label{eq:SNR3}
    \frac{{\cal{Q}}_3^2}{\sigma^2({\cal{Q}}_3)} &\sim & 
    \left( \frac{\bar{n}_1 \, \bar{n}_2}{\bar{n}_1 + \bar{n}_2}  \right)^2
    \to \left( \frac{\bar{n}_T}{4} \right)^2 \; ,
\end{eqnarray}
where, for two tracers, $\bar{n}_T = \bar{n}_1 + \bar{n}_2$, and the expressions on the right-hand sides result from using the optimal ``equipartition'' configuration that maximizes the SNR, namely $\bar{n}_1=\bar{n}_2=\bar{n}_T/2$.

The first limit, Eq. \eqref{eq:SNR1}, expresses cosmic variance: the accuracy in measurements of the matter power spectrum is fundamentally limited by the volume where we measure the density perturbations and by the width of the bandpower, even if we can count on arbitrarily large numbers of tracers to determine the density field inside that particular volume.
The second limit, Eq. \eqref{eq:SNR2}, means that ratios of power spectra of different tracers can be measured with an accuracy that is only limited by the numbers of tracers that we have, and scale with the number density \cite{abramo2013multitracer}.
Finally, the third limit, Eq. \eqref{eq:SNR3}, tells us that the irreducible degrees of freedom of the cross-correlations scale even faster with the tracer densities, compared with the ratios of tracers. Therefore, in the limit of high number of tracers, these degrees of freedom can be determined with extremely high accuracy, and are even less constrained by cosmic variance.

On the other hand, in the limit of small number densities, the situation is reversed. In that case the Fisher matrix of Eq. \eqref{eq:DiagFishPk} tells us that the SNR of the degrees of freedom scale as
$(\bar{n}_1 + \bar{n}_2)^2 $ for ${\cal{Q}}_1$,
as $\bar{n}_1 \, \bar{n}_2 $ for ${\cal{Q}}_2$,
and as $ \bar{n}_1^2 \, \bar{n}_2^2/(\bar{n}_1 + \bar{n}_2)^4 $ for ${\cal{Q}}_3$.
Therefore, in the limit of very sparse tracers it becomes increasingly difficult to measure the ratios of spectra, and even harder to determine the irreducible degrees of freedom in the cross-spectra.

These results put the issue of the cancellation of cosmic variance into a new light: some degrees of freedom profit even more from a denser sampling than others. 
Ratios of spectra (green lines in Fig. \ref{fig:SNR}), which allow us to measure, e.g., redshift-space distortions and primordial non-Gaussianities, as well as many parameters in the bias expansion \cite{2021arXiv210811363M}, start to become more interesting when the numbers of tracers are larger than $\bar{n} \gtrsim 10^{-4} \, h^{3} \, {\rm Mpc}^{-3}$, but they saturate at a SNR $\sim 1-2\%$ unless we have both tracers with number densities $\bar{n} \gtrsim 10^{-2} \, h^{3} \, {\rm Mpc}^{-3}$, which doesn't seem reasonable.
On the other hand, the independent degrees of freedom in the  cross-spectrum (red lines in Fig. \ref{fig:SNR}) start to become detectable for $\bar{n} \gtrsim 10^{-4} \, h^{3} \, {\rm Mpc}^{-3}$, but their accuracy grows even faster with the number density, such that we can reasonably achieve accuracies of $\sim 0.3\%$ when the two tracers have number densities of $\bar{n} \sim 10^{-3} \, h^{3} \, {\rm Mpc}^{-3}$.

Finally, we note that we were unable to determine an analytic expression that generalizes to $N$ tracers the independent degrees of freedom ${\cal{Q}}_i$ when we include the cross-spectra. Nevertheless, a numerical study of the eigenvalues of the Fisher matrix seems to indicate that the irreducible degrees of freedom in the cross-spectra indeed scale in the manner shown in Eq. \eqref{eq:SNR3}. We will return to this issue in a future paper.

\section{Conclusions}

This paper conveys two main results. Firstly, we showed how to use two summary statistics of the Fisher matrix in order to optimize the number of tracer species in a survey. 
We employed the following proxies for the total Fisher information: the determinant (or Figure of Merit), and the grand sum of the Fisher matrix.
By assuming a simple power-law relation between bias and number density we have shown that using either one of these two summary statistics, with or without the explicit inclusion of the independent degrees of freedom for the cross-spectra, the configuration that optimizes the total Fisher information is that in which the tracers are divided into equal samples: $\bar{n}_i = \bar{n}_T/N$, where $\bar{n}_T = \sum_{i=1}^N \bar{n}_i$ is the total number of tracers in the survey. 
Moreover, in general, the higher the total clustering SNR
${\cal{P}} = \sum_i \bar{n}_{i} P_{ii}$, the higher is the optimal number of tracers that one should employ in order to maximize that Fisher information.

We have also shown in this paper that the information in the irreducible degrees of freedom of the cross-spectra, expressed in terms of ${\cal{P}}_{ij}^2 - {\cal{P}}_{ii}{\cal{P}}_{jj}$, can be measured with an accuracy that is not limited by cosmic variance. 
This result is analogous to the well-known fact that ratios of spectra are partially immune to cosmic variance, however, while the accuracy of spectral ratios increase with $\sim \sqrt{\bar{n}_T}$, the accuracy of the degrees of freedom of the cross-spectra grow with $\sim \bar{n}_T$.
This means that physical parameters which are manifested in the combination ${\cal{P}}_{ij}^2 - {\cal{P}}_{ii}{\cal{P}}_{jj}$ can be measured, at least in principle, to exquisite precision.
In particular, stochastic and non-linear terms in the perturbative bias expansion, such as those shown in Eq. \eqref{eq:AutoCross}, can benefit from this windfall of the multi-tracer analysis.

\acknowledgments
We thank Henrique Rubira, Thiago Mergulh\~ao and Rodrigo Voivodic for useful comments. We also acknowledge the financial support of FAPESP (R.A.), CNPq (R.A \& I.L.T.) and CAPES (J.V.D.F.).

\nocite{*}
\bibliographystyle{JHEP}
\bibliography{stuff.bib}

\end{document}